\documentclass[reqno]{amsart}
\usepackage{amsmath}
\usepackage{cite}
\theoremstyle{definition}

\newtheorem{remark}{Remark}

\newcommand{\p}{\partial}

\newcommand{\dd}{{\rm d}}

\title[A unifying mechanical equation with applications] %Use the shortened version of the full title
      {A unifying mechanical equation with applications to non-holonomic constraints and\\ dissipative phenomena}

\author[E. Minguzzi]{}

 \email{ettore.minguzzi@unifi.it}

\begin{document}
\maketitle

% Enter the first author's name and address:
\centerline{\scshape E. Minguzzi}
\medskip
{\footnotesize
% please put the address of the first author
 \centerline{Dipartimento di Matematica e Informatica ``U. Dini''}
   \centerline{Universit\`a
degli Studi di Firenze, Via S. Marta 3,  I-50139 Firenze, Italy.}
   %\centerline{ Springfield, MO 65801-2604, USA}
} % Do not forget to end the {\footnotesize by the sign }

\bigskip

% The name of the associate editor will be entered by an editorial staff
% "Communicated by the associate editor name" is not needed for special issue.
 %\centerline{(Communicated by the associate editor name)}

%The abstract of your paper
\begin{abstract}
A  mechanical covariant equation is introduced which retains all the effectingness of the Lagrange equation while being able to describe, in a unified way, other phenomena including friction,  non-holonomic constraints and energy radiation (Lorentz-Abraham-Dirac force equation).
A quantization rule adapted to the dissipative degrees of freedom is proposed which does not pass through the variational formulation.
\end{abstract}

%The title of your section 1
\section{Introduction}

According to Lagrangian mechanics the motion of a mechanical  system subject to holonomic constraints is
described by the Lagrange equation
\begin{equation} \label{xod}
\frac{\dd }{\dd t} \frac{\p L}{\p \dot q^k}- \frac{\p L}{\p
q^k}=Q^{(np)}_k ,
\end{equation}
where $L=T-V$ is the Lagrangian and $Q_k^{(np)}$ is a generalized non-potential force.
The left-hand side admits a variational interpretation
%\[
%\delta \int L \,\dd t=-\int (\frac{\dd }{\dd t} \frac{\p L}{\p \dot q^k}- \frac{\p %L}{\p
%q^k})\,\delta q^k \,\dd t.
%\]
% The variational approach has
   which has proved extremely successful, well beyond the realm of analytical mechanics, and particularly in quantum field theory through the path integral formulation.

Still, already at the classical level the variational interpretation reveals some limitations as it does not seem able to unify other aspects of mechanics. This is mainly due to the right-hand side of Eq.\ (\ref{xod}) which does not really admit a variational formulation and which appears whenever the system is not isolated, but rather in contact with other unaccounted for degrees of freedom. These difficulties are particularly clear in the following applications\\
\begin{itemize}
\item[(i)] {\bf Friction forces}. The most effective way of dealing with mechanical problems involving friction dependent on the velocity is through the introduction of the Rayleigh dissipation function $R(t,q,\dot q)$.
    \begin{equation}\label{ray}
    \frac{\dd }{\dd t} \frac{\p L}{\p \dot q^k}- \frac{\p L}{\p
q^k}=-\frac{\p R}{\p \dot q^k}.
    \end{equation}
    Lurie has show \cite{minguzzi14f,lurie02} that this approach works very well also for friction forces non-linear in the velocities.

    The above equation is non-variational, indeed Bauer \cite{bauer31}
    showed that it is impossible to obtain a single linear dissipative equation of motion with constant coefficients from a variational principle.
        Many approaches have been proposed to overcome these difficulties and extend variational Lagrangian mechanics to the realm of dissipative forces. Bateman  \cite{bateman31} observed that the degrees of freedom can be doubled in such a way that, when half of them are removed from the Lagrange equations, one obtains dynamical equations exhibiting friction.
        A doubling mechanism related to complex conjugation was proposed by Dekker \cite{dekker81} and more recently a general framework relating the doubled variables with a
        Hamiltonian principle with modified boundary conditions has been proposed by Galley \cite{galley13,kuwahara13}.
        We mention Riewe \cite{riewe96} who proposed to build the Lagrangian from fractional derivatives. Finally,
        some strategies involving the replacement of the Lagrangian $L$ with $e^{\gamma t}L$ have also been proposed \cite{levicivita95,denman66}.

        These approaches have their how advantages, and clarify different aspects of the dissipative phenomena. However, the Rayleigh dissipation function approach is  the simplest and the most effective when it comes to applications. This is so because in mechanics we can construct this function from those of the different surfaces at contact taking advantage of some ready to use expressions, much as we do for the  kinetic energy    \cite{lurie02,minguzzi14f}.\\
        %
%        Most of these approaches work well only in some special cases, typically for friction linear in  the velocity and do not come close to the effectingness of the Rayleigh dissipation function approach when it comes to applications.
\item[(ii)] {\bf Holonomic and non-holonomic constraints}. Suppose that the mechanical system satisfies the following set of constraint equations
    \begin{equation} \label{all}
    f_\alpha(t,q,\dot q)=0, \qquad \alpha=1,\cdots, m
    \end{equation}
    which we assume to be independent in the sense that  the Jacobian $\p f_\alpha/\p \dot q^k$ has the maximum rank $m$. The holonomic constraints fall into this class, it suffices to differentiate them with respect to time.
    The accepted equations of motion  \cite{rund66,marle98} are given by (\ref{all}) and (repeated indices are summed)
    \begin{equation} \label{moa}
    \frac{\dd }{\dd t} \frac{\p L}{\p \dot q^k}- \frac{\p L}{\p
q^k}=\mu_\alpha \frac{\p f_\alpha}{\p \dot q^k} ,
\end{equation}
where $\mu_\alpha(t)$ are multipliers.

It is well known that the naive replacement $L\to L+\mu_\alpha f_\alpha$ in the variational principle does not return the above equations. This issue has long been discussed, the literature on non-holonomic mechanics being extensive \cite{bloch03,neimark04}. This observation  is also important for the quantization of non-holonomic systems as this operation becomes non-trivial due to the lack a variational formulation \cite{bloch08}.\\
\item[(iii)] {\bf Energy radiation}. The Lorentz-Abraham-Dirac force equation  (LAD)
describes the motion of a charged point particle. It takes into account the radiated  energy according to the Larmor power formula  (we use units such that $c=1$) $P_L=\frac{2}{3} \,q^2 \, {\bf a}^2$. The non-relativistic version is
    \begin{equation} \label{njx}
m {\bf a}=q ({\bf E}+{\bf v} \times {\bf B})+\frac{2}{3}\, q^2  \,  \dot {\bf a}
\end{equation}
while the relativistic version is (we use the Minkowski metric of signature $(-,+,\cdots,+)$)
\[
m a^\mu= q F^\mu_{\ \nu} u^\nu+ \frac{2}{3} \,q^2 \, \big( \frac{\dd a^\mu}{\dd \tau}-a^\alpha a_\alpha u^\mu \big),
\]
where $\tau$ is the proper time, $u^\mu$ is the covariant velocity and $a^\mu=\dd u^\mu/ \dd \tau$ is the covariant acceleration. There have been other proposal, like that by Landau-Lifshitz which can be considered a perturbative version of LAD which holds whenever the
term in the derivative of the acceleration (jerk) is small compared to the other terms.

Since this  equation is of third order the impossibility of obtaining it from a variational principle should be clear. Indeed, the usual Lagrange equation is of second order while if we try to include the accelerations in the Lagrangian $L(t,q,\dot q,\ddot q)$ we obtain an equation
\begin{equation} \label{aap}
\frac{\p L}{\p q^k}-\frac{\dd }{\dd t} \frac{\p L}{\p \dot q^k}+\frac{\dd^2}{\dd t^2} \frac{\p L}{\p \ddot q^k}=0
\end{equation}
which is of fourth order. Thus following a variational approach one can at most hope to recover the derivative of the LAD equation \cite{carati98} (it can be mentioned that the  equations of motion of spinning particles can be obtained from fourth order PDEs of variational origin \cite{riewe72,matsyuk11}).

In general, the increase in the differentiability order signals that some degrees of freedom have been ignored (recall that a system of differential equations at some order might reduce to a system of less equations at a higher order). These lost degrees of freedom might signal their existence through dissipative effects. In the LAD case, these degrees of freedom are due to the electromagnetic field, but there are also simpler mechanical systems where the same phenomena might take place, for instance in the mechanical phenomena of stick-slip friction. %\cite{varagnolo13}.
\end{itemize}

The previous examples show that for the treatment of some specific problems we have a set of working recipes rather than a unifying mathematical formalism. One would like to find a new formulation of mechanics which could unify the above solutions. It is quite clear that such unification cannot pass through the standard variational methods, that is, it cannot simply consist in adding suitable terms to the  Lagrangian.

A suggestion is provided by the key ingredient which makes Lagrangian mechanics so effective in the solution of mechanical problems. This is not the variational nature of the Lagrange equations, but rather their {\em covariance}. This is the key property that allows one to remove the constraint forces from the analysis. It turns out that though every variational equation is covariant, there are covariant equations which have no variational origin.

We now show how to unify mechanics using the simplest of these equations. The dynamical equation is
\begin{equation} \label{pwk}
 \frac{\p P}{\p \dot q^k}-2\frac{\dd }{\dd t}  \frac{\p
P}{\p \ddot q^k} =0 ,
\end{equation}
where $P(t,q,\dot q,\ddot q)$ has the dimension of power. We call it {\em power Lagrangian}. The left-hand side of (\ref{pwk}) is covariant in the sense that if $\tilde q=\tilde q(t,q)$, and $\tilde P$ is defined through $\tilde P(t,\tilde q,\dot{\tilde{q}},\ddot{\tilde{q}})=P (t,q,\dot q,\ddot q)$  then
\[
\frac{\p P}{\p \dot q^k}-2\frac{\dd }{\dd t}  \frac{\p
P}{\p {\ddot{q}}^k} =\Big(\frac{\p {\tilde P}}{\p \dot{\tilde{q}}^j}-2\frac{\dd }{\dd t}
\frac{\p {\tilde P}}{\p \ddot{\tilde{q}}^j} \Big) \frac{\p  \tilde{q}^j}{\p  q^k}.
\]

In order to prove its covariance first observe that
\begin{align}
\dot{\tilde{q}}^j&=\frac{\p \tilde{q}^j}{\p q^k}\, \dot q^k+\frac{\p \tilde q^j}{\p t}, \label{aod}\\
\ddot{\tilde{q}}^j&=\frac{\p \tilde q^j}{\p q^k}\, \ddot q^k+\frac{\p^2 \tilde q^j}{\p q^l\p q^k}\, \dot q^l \dot q^k+2 \frac{\p^2 \tilde q^j}{\p t\p q^k}\, \dot q^k+ \frac{\p^2 \tilde q^j}{\p t^2} \label{aos}
\end{align}
both $\dot{\tilde{q}}$ and $\ddot{\tilde{q}}$ depend on $\dot{q}$ thus
\begin{align*}
\frac{\p P}{\p \dot q^k}-2\frac{\dd }{\dd t}
\frac{\p P}{\p \ddot q^k} & =\frac{\p \tilde P}{\p \dot{\tilde{q}}^j} \frac{\p \dot{\tilde{q}}^j}{\p \dot q^k} +\frac{\p \tilde P}{\p \ddot{\tilde{q}}^j} \frac{\p \ddot{\tilde{q}}^j}{\p \dot q^k}-2\frac{\dd }{\dd t} \big( \frac{\p \tilde P}{\p \ddot{\tilde{q}}^j} \frac{\p \ddot{\tilde{q}}^j}{\p \ddot q^k}\big)\\
&=\frac{\p \tilde P}{\p \dot{\tilde{q}}^j} \frac{\p  \tilde{q}^j}{\p  q^k} +\frac{\p \tilde P}{\p \ddot{\tilde{q}}^j} \Big(2\frac{\p^2 \tilde q^j}{\p q^l\p q^k}\, \dot q^l +2 \frac{\p^2 \tilde q^j}{\p t\p q^k}\Big)-2\frac{\dd }{\dd t} \big( \frac{\p \tilde P}{\p \ddot{\tilde{q}}^j} \frac{\p  \tilde q^j}{\p  q^k}\big)\\
&=\Big(\frac{\p \tilde P}{\p \dot{\tilde{q}}^j}-2\frac{\dd }{\dd t}
\frac{\p \tilde P}{\p \ddot{\tilde{q}}^j} \Big) \frac{\p  \tilde q^j}{\p  q^k}.
\end{align*}
%this result proves the form invariance of the equations independently of the classical interpretation of $P$ as given in the previous section.

Observe that in the previous calculation we showed
\begin{equation} \label{clz}
\frac{\p P}{\p \ddot q^k}=\frac{\p \tilde P}{\p \ddot{\tilde{q}}^j} \frac{\p  \tilde q^j}{\p  q^k}
\end{equation}
This equation and (\ref{aos}) prove that the statement ``$P$ is a function independent of (or affine, quadratic on)  the acceleration'' is well posed as independent of the coordinate system.

It remains to specify  $P$. We give the prescription directly for the most general case which includes  constraints. Here $P$ must be augmented to include the multipliers as independent variables $P(t,q,\dot q,\ddot q, \lambda, \dot \lambda, \ddot \lambda)$. This is done in the most naive way as $P$ is given by
\begin{equation} \label{dec}
P=\frac{\dd L}{\dd t}-R+\dot \lambda_\alpha f_\alpha+P_E+Q^r_{k}\dot{q}^k ,
\end{equation}
where $P_E(t,q,\dot q,\ddot q)$ is a term non-linear in the generalized accelerations and $Q^r_k(t,q)$ is any other unmodeled generalized force. Function $P$ depends on the vector $\lambda_\alpha$ and its derivative only if there is some constraint.
From the easily checked identity
\begin{equation} \label{ail}
\frac{\p }{\p \dot q^k}\frac{\dd L}{\dd t}-2\frac{\dd }{\dd t} \Big(\frac{\p
}{\p \ddot q^k} \frac{\dd L}{\dd t}\Big)=\frac{\p L}{\p
q^k}-\frac{\dd }{\dd t} \frac{\p L}{\p \dot q^k}
\end{equation}
it is easy to verify that our equation implies Eqs.\ (\ref{xod}), (\ref{ray}), (\ref{all}) and (\ref{moa}) in their specific cases. Observe that (\ref{pwk}) is really of second order unless $P_E\ne 0$ which is therefore the radiating term.
The letter $E$ stands for {\em emission}.

Roughly speaking (\ref{pwk}) establishes an equilibrium of forces starting from their representative  power Lagrangian functions. In order from left to right these forces are: the inertial force and the conservative force, the force due to friction, the force due to the additional constraints, the force due to energy radiation and a final unmodeled force.

\section{A variational interpretation}

Equation (\ref{pwk}) has a kind of variational origin, indeed introduced the integral
\[
I[a,b]=\int_a^b \dd t \int_a^t P(u,q(u),\dot q(u),\ddot q(u)) \, \dd u ,
\]
through successive integration by parts we obtain
\begin{align*}
\delta I[a,b]&=\frac{\p P}{\p \ddot q^k} \, \delta q^k \Big|^b_a-\Big(\frac{\p P}{\p \ddot q^k} \, \delta \dot q^k+\frac{\p P}{\p \dot q^k} \, \delta  q^k- \frac{\dd }{\dd t}\frac{\p P}{\p \ddot q^k} \, \delta  q^k\Big)  \Big|_{t=a} (b-a)\\
&\quad +\int_a^b X_k(t)\,\delta q^k(t) \, \dd t +\int_a^b \dd t \int_a^t Y_k(u) \, \delta q^k(u) \dd u ,
\end{align*}
where
\begin{align*}
X_k(t)&=\frac{\p P}{\p \dot q^k}-2\frac{\dd }{\dd t}
\frac{\p P}{\p \ddot q^k},\\
Y_k(t)&=\frac{\p P}{\p q^k}-\frac{\dd }{\dd t} \frac{\p P}{\p \dot q^k}+\frac{\dd^2}{\dd t^2} \frac{\p P}{\p \ddot q^k}.
\end{align*}
The variational origin of these expressions makes them covariant as it is well known for the latter and as we checked for the former. Let us set $\delta q^k=\epsilon v^k \eta_\epsilon(t-\bar t)$ where $v^k$ is a fixed vector, $\eta_\epsilon$ is a mollifier and $\bar t\in (a,b)$. Since for $\epsilon\to 0$, $\eta_\epsilon$ converges to a Dirac's delta, the variation reduces to the leading order in $\epsilon$ to
\[
\delta I[a,b]=\epsilon v^k [X_k(\bar t)+Y_k(\bar t) (b-\bar t)].
\]
This expression has the drawback of depending on  $b$, while we want to get an equation which is independent of the extremes of the variational integral. If we impose that the first variation  vanishes for any interval $[a,b]$ then we get that both $X_k=0$ and $Y_k=0$ which is a too restrictive condition. We can recover just $X_k=0$ imposing that for small $\vert b-a\vert$ the first variation vanishes.
% in other words that at the leading order the expansion in the variational field and the interval length should go as $o(\epsilon)+O(\epsilon \vert b-a\vert)$. Thus $X_k=0$ over a curve $q(t)$ iff this curve is locally a stationary point of $I[a,b]$ for any infinitesimally small interval $[a,b]$.

In other words we have the identity
\[
\lim_{a,b \to t} \frac{\delta I[a,b]}{\delta q^k}=\frac{\p P}{\p \dot q^k}-2\frac{\dd }{\dd t}  \frac{\p
P}{\p \ddot q^k},
\]
thus every solution of (\ref{pwk}) can be regarded as a stationary point of a local  action involving a double integral. Without the limit $a,b\to t$ the variation would give a differential equation involving further terms dependent on the value of the boundary points $a,b$. In principle this feature could be exploited to model non-local or history dependent dissipative forces. This direction of investigation will not be pursed in this work.

\begin{remark} \label{rei}
An anonymous referee has  pointed out
%helped us to trace
some  history on the geometry of (\ref{pwk}). The covariance of the left hand side of (\ref{pwk}) appeared  in general studies by Craig \cite{craig35} and Synge \cite[Eq.\ (3.10)]{synge35}, where the function $P$ was not given a particular meaning. Bucataru and Miron \cite{bucataru09} have shown that this covector can also be obtained through a special variation of $\int P \dd t$ which involves only the velocities and the accelerations. If $P$ has non-singular Hessian with respect to the accelerations (a case to be considered in the next sections), then (\ref{pwk}) defines a third order differential equation, or a semispray.
These works were concerned with the geometrical aspects of covariant equations.
The possibility of recovering Lagrangian mechanics, including dissipation, constraints and energy radiation, passing from (\ref{pwk}) and (\ref{ail}), does not seem to have been previously noticed.
\end{remark}

\section{The LAD equation}

Since (\ref{pwk}) unifies problems (i) and (ii) in a common formalism and since when $P$ is non-linear in the acceleration it is of third order, one might ask whether it is able to comprise  (iii) for which a definite dynamics induced by energy radiation has been identified. The answer is affirmative.
\begin{itemize}
\item {\bf Non-relativistic version}.
%Let us introduce the electromagnetic potentials through
%\begin{align*}
%{\bf E}&=-\frac{1}{c} \frac{\p {\bf A}}{\p t}- \nabla \Phi, \qquad
%{\bf B}=\nabla \times {\bf A}.
%\end{align*}
Let $(\Phi,{\bf A})$ denote the electromagnetic potential.
It is sufficient to choose
\begin{align*}
L&=\frac{m}{2} \,{\bf v}^2-q \Phi+ q\, {\bf v}\cdot {\bf A},\\
P_E&= -\frac{1}{4} \,P_L=-\frac{q^2}{6}  \, {\bf a}^2
\end{align*}
where the former expression is the usual non-relativistic Lagrangian. The equation (\ref{pwk}) for Cartesian coordinates is the Lorentz-Abraham force.\\

\item {\bf Relativistic version}.
Let us introduce the electromagnetic potential through $F_{\mu \nu}=\p_\mu A_\nu-\p_\nu A_\mu$.
Let $x^\mu(s)$ be a curve and let us  denote differentiation with respect to $s$ with a prime.
We set
\begin{align}
L=&-m\sqrt{-x'^\alpha x'_\alpha}+qA_\alpha x'^\alpha, \\
P_E=&-\frac{q^2}{6}\, a_\mu a^\mu , \label{sed}
\end{align}
 where $a_\mu$ is a short-hand for the covariant acceleration
 \begin{align} \label{acc}
 a_\mu=\gamma (\gamma x'_\mu)'= \gamma^2 x''_\mu+\gamma^4 (x'^\beta x''_\beta) \,x'_\mu,
 \end{align}
with $\gamma=(-x'^\beta x'_\beta)^{-1/2}$. Observe that  $a_\mu x'^\mu=0$.
 %We can assume that $q\ne 0$ since $L$ is the Lagrangian for the Lorentz force equation.
Equation (\ref{pwk}) gives
\begin{equation}
\begin{aligned}
ma_\mu &=q F_{\mu \nu} \gamma x'^\nu+\frac{2}{3} \, q^2 \gamma^2 \Big\{\gamma \frac{\dd a_\mu}{\dd s}  -(a_\beta a^\beta)\, \gamma x'_\mu+\frac{3}{2}\, \gamma^3 (x'^\beta x''_\beta) \, a_\mu\Big\} . \!\!\!\!\!\!\!\!\!{}\label{ddp}
\end{aligned}
\end{equation}
 %Contraction with  $x'^\mu$ gives $a_\mu x''^\mu =a_\beta a^\beta$. Observe from (\ref{acc}) that $a_\mu x'^\mu=0$ thus multiplying Eq.\ (\ref{acc}) by $a^\mu$
%\[
%a_\mu a^\mu=\frac{x''_\mu a^\mu}{(-x'^\alpha x'_\alpha)}=\frac{a_\mu a^\mu}{(-x'^\alpha x'_\alpha)}.
%\]
%Thus in the open $s$-interval $U\subset \mathbb{R}$ where $ a_\mu  a^\mu \ne 0$, we have that $-x'^\alpha x'_\alpha=1=\gamma$ thus $s$ is the proper time, which implies $x'^\beta x''_\beta=0$ thus $x'^\mu$ is timelike and $a^\mu$ is spacelike. By continuity $-x'^\alpha x'_\alpha=1$ also on $\bar{U}$ and on the open set $\bar{U}^C$ we have $a_\mu  a^\mu =0$ so that as $P_E=0$ the equation coincides with the Lorentz force equation thus it preserves the causal character of the tangent vector. The normalization $-x'^\alpha x'_\alpha=1$ on its boundary shows that it is timelike. Thus $x'^\alpha$ is timelike everywhere, hence as $a_\mu x'^\mu=0$, $a_\mu$ is spacelike or zero everywhere, thus
%From this equation it is possible to show that $x'^\alpha$ is timelike everywhere, $ a_\mu  a^\mu \ne 0$ is equivalent to $a_\mu\ne 0$, and wherever this condition holds $s$ is the proper time so that $x'^\beta x''_\beta=0$ and $\gamma=1$. This observation shows that the last term on Eq.\ (\ref{ddp}) vanishes, thus giving the Lorentz-Abraham-Dirac force once the curve is parametrized with respect to proper time (the reparametrization being necessary only where there is no acceleration).
Over a proper time parametrized curve ($\gamma=1$) this equation coincides with the  Lorentz-Abraham-Dirac force equation, thus whether the (only) solution is a solution of LAD depends on the initial conditions which have to satisfy $x'^\beta x'_{\beta}=-1$, $x'^\beta x''_{\beta}=0$.

There is also another relativistic treatment which  fixes the  parametrization to proper time.
 Let us write our equations with $t=\tau$ (without invoking in advance any special meaning for this label), and let us denote $x^\mu=q^\mu$, $u^\mu=\dot q^\mu$, $a^\mu=\ddot q^\mu$. We set
\begin{align}
L&=\frac{m}{2} \, u^\alpha u_\alpha+q A_\alpha u^\alpha, \\
P_E&=-\frac{q^2}{6} \,  \frac{a^\alpha a_\alpha}{(u^\beta u_\beta)^2} \label{sec}.
\end{align}
Then imposing the initial condition $u^\alpha u_\alpha=-1$, $a^\mu u_\mu=0$, Eq.\ (\ref{pwk}) after some manipulations gives the LAD equation.
%
%Let us check more closely this derivation\footnote{This calculation does not appear in the published version, and is included here for the author reference.}. Equation (\ref{pwk}) reads
%\begin{align*}
%m a_\mu=q F_{\mu \nu} u^\nu+ \frac{2 }{3} \frac{q^2}{(u^\beta u_\beta)^2}\, \left( \frac{d a_\mu}{\dd \tau}-(a^\alpha a_\alpha) \frac{u_\mu}{-u^\beta u_\beta} -4 \frac{a_\alpha u^\alpha}{u^\beta u_\beta} \,a_\mu \right)
%\end{align*}
%so multiplying by $u^\mu$ and setting $z=u^\mu u_\mu$
%\[
%\frac{1}{2} \,m  \frac{\dd z }{\dd \tau}=\frac{2 }{3} \,q^2 \Big[\frac{1}{2z^2}  \frac{\dd^2 z }{\dd \tau^2}-\frac{1}{z^3}(\frac{\dd z }{\dd \tau})^2\Big] ,
%\]
%which can be rewritten
%\[
%\frac{\dd }{\dd \tau} \Big[m z-(\frac{2 }{3} q^2)\frac{1}{z^2} \frac{\dd z}{\dd \tau}\Big]=0.
%\]
%Under the initial condition $z=u^\alpha u_\alpha=-1$ and $\frac{\dd z}{\dd \tau}=2 u^\alpha a_\alpha=0$ it is equivalent to
%\[
%k\frac{\dd z}{\dd \tau}= (z+1) z^2, \qquad k= \frac{2 q^2}{3 m} ,
%\]
%which has a Lipschitz right-hand side, thus it has a unique solution.
%We conclude that the unique solution  is $z=u^\alpha u_\alpha=-1$ which proves that the
%last term of dynamical equation vanishes, thus giving the Abraham-Lorentz-Dirac force.
\end{itemize}
%It can be observed that in the first relativistic approach the initial condition of second order are important in order to recover the LAD equation. This initial condition are not important in the second relativistic approach.

Finally, it is interesting to investigate possible first integrals of (\ref{pwk}).
Recalling the  decomposition (\ref{dec}),  let us define $F$ so that $P=\frac{\dd L}{\dd t}+F$ and
\[
U=2 \frac{\p F}{\p \ddot q^j}\,\dot q^j+ \frac{\p L}{\p \dot q^j}\,\dot q^j-L .
\]
%Let us define
%\[
%U=2 \frac{\p P}{\p \ddot q^j}\,\dot q^j
%\]
Multiplying (\ref{pwk}) by $\dot{q}^k$ we obtain
after some manipulations
\begin{equation} \label{spp}
\frac{\dd U}{\dd t}=\frac{\p F}{\p \dot q^k} \, \dot q^k+2 \frac{\p F}{\p \ddot q^k} \, \ddot q^k-\frac{\p L}{\p t} .
\end{equation}
%Using the decomposition (\ref{dec}) one can easily show that for $P=\frac{\dd L}{\dd t}$ we have $U=\frac{\p L}{\p \dot q^j} \,\dot q^j-L$ and
%\[
%\frac{\p P}{\p \dot q^k} \, \dot q^k+2 \frac{\p P}{\p \ddot q^k} \, \ddot q^k=-\frac{\p L}{\p t}.
%\]
  If $F=0$ the function $U$ is the usual Hamiltonian which is conserved whenever the Lagrangian is independent of time. Other invariants are obtained whenever $F$ includes a   radiative term $P_E$ that satisfies the positive homogeneity property
\begin{equation} \label{hom}
P_E(t,q,s\dot q,s^2\ddot q)=P_E(t,q,\dot q,\ddot q), \qquad \forall s>0
\end{equation}
for which the first two terms on the right-hand side of (\ref{spp}) vanish. For instance this type of symmetry is found in the study of the LAD equation, see Eq.\ (\ref{sec}),
%or (\ref{sed})-(\ref{acc}),
where the invariant is connected with the normalization $u^\mu u_\mu=-1$.
%The analogous condition for the constraint term reveals that it must be linear in the velocities to be conservative.
 %while that for the friction force shows that the Rayleigh function must depend only on the direction of the velocity. If we assume that the friction force is collinear with the velocity then it must vanish.

We now work out  a simple exercise which shows how effective  Eq.\ (\ref{pwk}) is in applications.\\

\subsection{A constrained electrically charged particle}
%\noindent {\em A constrained electrically charged particle.}\\
Let us consider a particle of mass $m$ and charge $q$ constrained to move on a horizontal parabola $y=\frac{b}{2}\, x^2$. Let us suppose that it undergoes Coulomb friction with coefficient $\mu$ and that emits radiation according to the Lorentz-Abraham force.
The velocity and acceleration read
\begin{align*}
  v^2&=\dot x^2(1+b^2 x^2)\\
  a^2&=\ddot x^2(1+b^2 x^2)+b^2\dot x^4+ 2b^2x \dot x^2 \ddot x .
\end{align*}
The contact force $N$ between the constraint and the point particle is $N=v^2/r$ where
\[
r^{-1}= y''/(1+y'^2)^{3/2}=b/(1+b^2x^2)^{3/2}
\]
is the  curvature. Since the Rayleigh dissipation function for Coulomb friction is $R=\mu N v$ (see \cite{lurie02,minguzzi14f}) we obtain
\begin{align*}
L&=\frac{1}{2} \,m (1+b^2 x^2 ) \dot x^2,\\
R&= \mu b\vert\dot x\vert^3\\
P_E&=-\frac{q^2}{6 } (\ddot x^2(1+b^2 x^2)+b^2\dot x^4+ 2b^2x \dot x^2 \ddot x),
\end{align*}
from which we obtain the dynamical equation (\ref{pwk})
\begin{align*}
m\{(1+b^2 x^2)\ddot x&+2 b^2 x \dot x^2\}=-3\mu b \dot x^2 \textrm{sgn}(\dot x)
 +\frac{2 }{3 }\, q^2\big\{3 b^2 x \dot x \ddot x+(1+b^2 x^2)\dddot x\big\}.
\end{align*}
Observe that one advantage  of our method stands on the possibility of using any coordinate system adapted to the constraints even in problems that involve dissipation due to friction or energy radiation. For an arbitrary charge distribution the radiating term $P_E$ can be deduced from the power radiated at infinity. Of course, it is not the sum of the Larmor powers of the single charges. For instance,  an electric dipole $p$ rotating with angular velocity $\omega$ emits radiation with power $P'=\frac{2 p^2 \omega^4}{3 }$ (cf.\ \cite[Chap.\ 9]{jackson75}). Thus an argument similar to that used to support the Abraham-Lorentz force shows that on the dipole acts a momentum ${\bf M}= - \frac{2 p^2 \omega^2}{3 } \, \boldsymbol{\omega}$, indeed ${\bf M}\cdot \boldsymbol{\omega}=-P'$. This friction enters  $R$ rather than $P_E$.\\

%The equation is of third order but it reduces to a second order equation whenever  the power Lagrangian is linear in the accelerations.

\section{A quantization of the dissipative degrees of freedom}
We end the paper with some speculations on the quantization of the theory. It is clear that we cannot apply the usual canonical quantization scheme since we don't have  at our disposal a Hamiltonian, this function being available only in absence of dissipative effects.
The easily checked identity
\[
\frac{\p }{\p \ddot{q}^k}\frac{\dd L}{\dd t}=\frac{\p L}{\p \dot q^k}
\]
 suggests to define, more generally, the generalized momenta  as $p_k={\p P}/{\p \ddot{q}^k}$.  Suppose for simplicity that all the degrees of freedom are radiative, namely that $P$ is non-linear in $\ddot{q}$ in the sense that the matrix ${\p^2 P}/{\p \ddot{q}^j\p \ddot{q}^k}$ is non-degenerate.
 Then the dependence on $\ddot q$   can be removed through a Legendre transformation by introducing the function $W(t,q,\dot q,p)=p_k \ddot q^k-P$, which allows us to rewrite the dynamical equations in the new coordinates as follows
\begin{align*}
\frac{\dd \dot q^k}{\dd t}&=\frac{\p W}{\p p_k},
 &\frac{\dd p_k}{\dd t} &=-\frac{1}{2}\frac{\p W}{\p \dot q^k},
\end{align*}
to which we can add
\begin{align*}
-\frac{\p P}{\p q^k}&=\frac{\p W}{\p q^k},
&- \frac{\p P}{\p t} &= \frac{\p W}{\p t} .
\end{align*}
This formulation in terms of first order PDEs is reminiscent of the Hamiltonian formulation of analytical mechanics. Nevertheless, these equations do not define a symplectic structure due to the $1/2$ factor on the second equation.

We now give a quantization rule inspired by the standard quantization. The usual quantization involves non-dissipative degrees of freedom and leads to the quantization of energy for some isolated systems. Here we proceed formally to a different quantization rule, adapted to the dissipative degrees of freedom and which might give the quantization of dissipative power.

To start with let us observe that the usual commutation rule for the non-dissipative degrees $[q^k,p_j]=i \hbar \delta^k_j$ is imposed between a variable ($q$) whose time derivative ($\dot q$) is conjugate in the Legendre sense to the other ($p$). So if we do the same here we have to impose an analogous commutation relation between a variable ($\dot q$) whose time derivative ($\ddot q$) is conjugate in the Legendre sense to the other ($p$). Thus we have to impose
\[
[\dot q^k,p_j]=i \Lambda\hbar \delta^k_j
\]
where $\Lambda$ is a fundamental constant with  the dimension of a frequency. The introduction of this new fundamental constant shows that this is really a new type of quantization. We are essentially introducing a quantization scheme for those degrees of freedom which, being radiative, are not usually included in the standard quantization. Whether it makes sense to proceed in this way should be a matter of further theoretical and experimental investigation. As it happens for the Dirac quantization rule here our proposed quantization is supposed to hold only for a Cartesian choice of the coordinates $\{q^k\}$.

As an application, let us consider a one dimensional bremsstrahlung phenomena, namely a  particle
of mass $m$ and charge $q$ subject to a viscous friction $-\gamma v$ on a one dimensional space $\mathbb{R}$, $v=\dot x$.
    We have
\[
P=m a v-\frac{1}{2}\, \gamma v^2-\frac{1}{2} \, \mu \, a^2
\]
where $\mu$ is a radiative coefficient proportional to $q^2$. Thus $p=m v-\mu a$ and
\[
W=p a-P=\frac{1}{2}\, \gamma { v}^2-\frac{1}{2} \,\mu { a}^2 %=\frac{1}{2}\, \gamma { v}^2-\frac{1}{2 \mu} \, ({ p}-m{ v})^2 .
\]
Let us consider the case in which $m$ is negligible, and hence the radiative force cannot be treated perturbatively. We can place $p=-\mu a$ and so the function $W$ has the structure of an inverted harmonic oscillator
\[
H:=(\frac{\gamma \hbar^2 \Lambda^2}{2})^{-1}W=-\frac{\p^2}{\p p^2}-\omega^2 p^2, \qquad \omega=\frac{1}{\sqrt{\gamma \mu}} \frac{1}{ \hbar \Lambda}.
\]
Adapting the calculations of \cite{yuce06} for the inverted oscillator we find a family of eigenstates for which $p$ (read the acceleration) is bounded, the bound increasing exponentially with time, and for which the difference between the power lost due to friction and that lost due to radiation  is quantized
\[
W_n=\frac{3}{2} \,{\gamma \hbar^2 \Lambda^2} \,e^{-4\omega t} \frac{\pi^2 n^2}{p_0^2} ,
\]
where $p_0=-\mu a_0 $ and $a_0$ is the initial acceleration. With time the maximum acceleration grows exponentially and the  quantas become smaller.

%As far as we know this is probably a unique model which predicts a quantization of power. Of course, it
This model relies on some speculative hypothesis introduced so as to extend the usual quantization rules to the dissipative degrees of freedom.

%\section*{Acknowledgments}

% You may incorporate your references as follows in your main tex file.
% Using BibTex is not recommended but can be handled.

\section{Conclusions}

We have presented a  covariant equation which retains all the effectiveness of the Euler-Lagrange equation, and might possibly be placed at the foundations of analytical mechanics.
 %could be placed at the foundations of analytical mechanics.
 Dissipation, energy radiation or nonholonomic constraints appear as specific contributions to the power Lagrangian, so the new formalism is able to unify otherwise unrelated solutions to different  problems. In the last section we have speculated on a possible quantization rule for the dissipative degrees of freedom, which we inferred from an analogy with the Dirac quantization rule. Its application suggests that the difference in the dissipated powers due to friction or radiation could be quantized.

\section*{Acknowledgments}   I thank an anonymous referee for mentioning previous work on the geometry of Equation (7), cf.\ Remark \ref{rei}.

%\bibliography{../../bibliografie/simultaneity,../../bibliografie/libri,../../bibliografie/miei,../../bibliografie/mieiPreprints,../../bibliografie/mieiProceedings}
%\bibliographystyle{AIMS}

%\medskip
% The data information below will be filled by AIMS editorial staff
%Received xxxx 20xx; revised xxxx 20xx.
%\medskip

\def\cprime{$'$}
\providecommand{\href}[2]{#2}
\providecommand{\arxiv}[1]{\href{http://arxiv.org/abs/#1}{arXiv:#1}}
\providecommand{\url}[1]{\texttt{#1}}
\providecommand{\urlprefix}{URL }

\end{document}